\documentclass[prc,unsortedaddress,reprint,amsmath,amsfonts,amssymb,showpacs,floatfix,nofootinbib]{revtex4-1}
\usepackage{graphicx}
\usepackage{multirow}
\usepackage{amsmath}
\usepackage{relsize}
\usepackage{longtable}
\usepackage{color}
\begin{document}

\title{Excited-state one-neutron halo nuclei within a parallel momentum distribution analysis}

\author{Shubhchintak}
\email{shub.shubhchintak@tamuc.edu}

\affiliation{Department of Physics and Astronomy, Texas A$\&$M University-Commerce, Commerce, TX 75429, USA}

\date{\today}

\begin{abstract}
Using a fully quantum mechanical post-form finite range distorted wave Born approximation theory of Coulomb breakup, I study the parallel momentum distribution of the core in the Coulomb breakup of suggested excited-state one-neutron halo nuclei considered in their different bound excited states. 
Narrow momentum distributions obtained in the present calculations for some cases indicate the possibilities of the excited-state halo structure in the nuclei under consideration and therefore favor the previous predictions.

\end{abstract}

\maketitle

\section {Introduction}
In the past few decades, with advances in radioactive ion beam facilities, it has become possible to explore the nuclei closer to the drip line. With this progress, some interesting structures have been observed in some nuclei, where a central core remains surrounded by the valence nucleon(s) forming a ``halo". These halo nuclei are characterized by having a long low density tail of loosely bound valence nucleon(s). So far, several one and two-nucleon halo nuclei have been observed in the low-mass region and some have also been suggested in the medium-mass region in or near the island of inversion \cite{jensen2004,tanihata13,nakamura1,shub1,Kobayashi,shub2}. These nuclei are found to have several different properties as compared to their stable isotopes. They exhibit a strong cluster structure of a core plus one or two nucleons. Halo nuclei generally have small one- or two-nucleon binding energy and low angular momentum ($\ell$) of the valence nucleon(s), preferably $s$ or $p$ wave. Because of the low value of $\ell$ (0 or 1) the centrifugal barrier causes almost no hindrance to the valence nucleon(s) and they can tunnel outside the classically allowed region. This results in enhancement in the root mean square (rms) radius of the halo nuclei even beyond the range of nuclear forces. As compared to neutron halo nuclei the formation of proton halo is less probable because of the Coulomb barrier. The two-nucleon halo nuclei, exhibit the {\it borromean} structure, a three-cluster system in which none of the two-body system is bound but the three-body system is bound \cite{zhukov}.        

Other features of halo nuclei include their large interaction or reaction cross sections, soft $E1$ excitations, and narrow momentum distributions. The first confirmation about the large radii of halo nuclei was obtained by Tanihata {\it et al.} \cite{tanihata} in the measurement of interaction cross sections of Li isotopes. Interestingly, $^{11}$Li was found to have large interaction cross sections as well as a large radius ($>$ $r_0$A$^{1/3}$) as compared to the other Li isotopes and now it is a well-established two-neutron halo nucleus. In the present study I consider only the one-neutron halo nuclei. The best example of a one-neutron halo nucleus is $^{11}$Be, which has an intruder configuration \cite{talmi} where the valence neutron occupies the 2$s_{1/2}$ orbital instead of the 1$p_{1/2}$ orbital. The other examples are $^{15}$C, $^{19}$C, and two recently suggested halo nuclei in the island of inversion, $^{31}$Ne and $^{37}$Mg.  

To date, all the well-established halo nuclei have been observed in their ground states except $^{17}$F, which has a halo structure in its first $1/2^+$ excited state \cite{morlock}. In Ref. \cite{szuki91}, the question about the possibility of a halo structure in the excited state of stable nuclei was raised. In fact, because of their short life time, it is difficult to perform measurements on the excited states. However, as mentioned in Ref. \cite{riisager}, the important information about the excited states can be extracted by measuring the electromagnetic transitions involving these states. In Ref. \cite{kobayashi2}, the recoil distance transmission method has been reported to measure the interaction cross sections of excited states. One can also determine the radii of the excited states using indirect measurements like the modified diffraction model \cite{danilov} and asymptotic normalization coefficients methods \cite{liu,lin,belyaeva}.
Some recent studies report about the halo structures in loosely bounded excited states of some nuclei \cite{liu,lin,belyaeva,otsuka1,otsuka,khalili,romero,arai,li,demyanova}. These include the 1/2$^-$(E$_x = 0.320$ MeV) state of $^{11}$Be \cite{belyaeva}, the 2$^-$ (E$_x = 1.673$ MeV) and $1^-$ (E$_x = 2.620$ MeV) excited states of $^{12}$B \cite{liu}, the 1/2$^+$ (E$_x = 3.089$ MeV) state of $^{13}$C \cite{otsuka,liu}, the 2$^-$ (E$_x = 6.263$ MeV) state of $^{10}$Be \cite{khalili}, and the first $1/2^+$ (E$_x = 1.684$ MeV) excited state of $^{9}$Be \cite{demyanova} (unbound state), which are predicted to have one-neutron halo structures.
%However, like the well established halo nuclei, there is a need to find more parameters in support of these studies.
% All the above cases are suggested to have halo structure because of the large spatial extension found in each case.  

In this paper, using a fully quantum mechanical theory of Coulomb breakup \cite{shub1,rc}, I study the parallel momentum distribution (PMD) of a charged fragment in the Coulomb breakup of suggested excited-state halo nuclei on a heavy target. Although the Coulomb breakup experiments in the excited states are difficult to date, in theory such a study can be made to do some predictions which of course need verifications from experiments.
In fact, experiments of the type reported in Refs. \cite{fleurot, kobayashi2} could be useful in this direction. For present calculations I use the post-form finite range distorted wave Born approximation (FRDWBA) theory of Coulomb breakup, which includes all order electromagnetic interactions between the fragments and the target \cite{rc}. It also includes the breakup contributions from the entire non-resonant continuum corresponding to all multipolarities and avoids the uncertainties associated with multipole distributions. The only input needed is the projectile ground state wavefunction. Previously, this theory has also been used in the study of ground state neutron halo nuclei in the light-mass as well as in the light-medium-mass region \cite{shub1,shub2,rc} and calculations were found in agreement with the data. Here, I use this theory to look for the possibility of a halo structure in the excited state of some nuclei which could be of interest for future experiments.   
It is well known that the full width at half maxima (FWHM) of the well-established halo nuclei (ground state) like $^{11}$Be and $^{19}$C is around 44 MeV/$c$ and is around 140 MeV/$c$ for the case of stable nuclei \cite{kelley,sauvan}. Therefore, I follow this criteria in the present study to predict a halo structure in the excited state of a nucleus.   

The paper is organized in the following way. In Sec. II, I give a brief formalism of the PMD in Coulomb breakup and then I discuss my results in Sec. III. Finally, in Sec. IV, I present the conclusions.  
%%%%%%%%%%%%%%%%%%%%%%%%%%%%%%%%%%%%%%%%%%%%%%%%%%%%%%%%%%%%%%%%%%%%%
\section{Formalism}
I consider the elastic breakup of a projectile $a$ [consisting of two clusters $b$ (core) and valence neutron $n$], in the Coulomb filed of target $t$ via the reaction $a + t \rightarrow b + n + t$. The Jacobi coordinate systems used is shown in Fig. 1.

\begin{figure}[ht]
\centering
\includegraphics[scale=0.55,trim={180 210 0 0cm},clip]{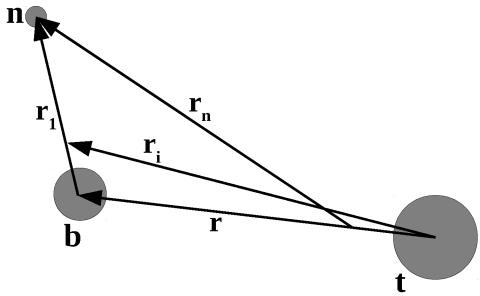}
\caption{The three-body Jacobi coordinate system.}
\label{Fig1}
\end{figure}

The position vectors in Fig. 1 are related to each other by the following relations:
\begin{eqnarray}
{\bf r} = {\bf r}_i -\alpha{\bf r}_1, \,\,\,\,\,\,\,\,\,  {\bf r}_n = \gamma{\bf r}_1 + \delta{\bf r}_i, \label{a1}
\end{eqnarray} 
where the mass factors $\alpha$, $\gamma$, and $\delta$ are given by
\begin{eqnarray}
\alpha=\frac{m_n}{m_n+m_b}, \,\,\,\,\, \delta = \frac{m_t}{m_b+m_t}, \,\,\,\,\, \gamma = (1-\alpha\,\delta),  \label{a2}
\end{eqnarray}
with $m_n$, $m_b$, and $m_t$ being the masses of fragments $n$, $b$, and $t$, respectively. 

Following Ref. \cite{shub3}, the PMD of the charged fragment $b$ in the present theory can be written as
\begin{eqnarray}
\frac{d\sigma}{dp_z}&=& 2\pi \, \int d\Omega_n\,dp_x\,dp_y\,\frac{m_b\,p_b}{\hbar\,v_a}\,\rho(E_b,\Omega_b,\Omega_n)\nonumber\\
&\times& \sum_{\ell\,m}\frac{|\beta_{\ell\,m}|^2}{(2\ell+1)}, \label{a3}
\end{eqnarray}
where $p_b$ is the momentum of core $b$ with $p_x$, $p_y$, and $p_z$ as its $x$, $y$, and $z$ components, respectively. $\rho(E_b, \Omega_b, \Omega_n)$ is the three-body phase-space factor in the final channel and $v_a$ is the $a-t$ relative velocity in the initial channel. $\ell$ and $m$ are the relative angular momentum between the constituents of the projectile and its projection, respectively. 

$\beta_{\ell\,m}$ in Eq. (\ref{a3}) is the reduced transition amplitude and is given by 
\begin{eqnarray}
\beta_{\ell\,m}({\bf q}_b,\,{\bf q}_n;{\bf q}_a) &=&\int\int d{\bf r}_1\,d{\bf r}_i \, \chi_b^{(-)*}({\bf q}_b,{\bf r})\, e^{-i{\bf q}_n.{\bf r}_n}\nonumber\\ 
&\times& V_{bn}({\bf r}_1) \phi_a^{\ell m}({\bf r}_1)\, \chi_a^{(+)}({\bf q}_a,{\bf r}_i), \label{a4}
\end{eqnarray}
where $V_{bn}$ is the interaction between the core $b$ and the neutron and ${\bf q}_j$ is the Jacobi wave vector of particle ``j". 
$\chi_b^{(-)}$ and $\chi_a^{(+)}$  are Coulomb-distorted waves for the relative motion of $b$ and the center of mass (c.m.) of $a$ with respect to the target $t$ with incoming and outgoing wave boundary conditions, respectively. $\phi_a^{\ell m}({\bf r}_1)$ is the bound-state wavefunction of the projectile. 

As explained in Ref. \cite{rc}, Eq. (\ref{a4}), which is a six-dimensional integral, can further be simplified using the local momentum approximation under which it splits into the product of two three dimensional integrals, as
\begin{eqnarray}
&\beta_{\ell\,m}({\bf q}_b,\,{\bf q}_n;{\bf q}_a) =\langle e^{i(\gamma\,{\bf q}_n-\alpha\,{\bf K}).{\bf r}_1}|V_{bn}|\phi_{a}^{\ell\, m}({\bf r}_1) \rangle \nonumber\\
&\times \langle \chi_b^{(-)}({\bf q}_b,\,{\bf r}_i)\,e^{i\delta\,{\bf q}_n.{\bf r}_i}|\chi_a^{(+)}({\bf q}_a,\,{\bf r}_i)\rangle, \label{a5}
\end{eqnarray}
where, the first integral contains the structure information of the projectile and is called the structure part, whereas the second integral is called the dynamics part and can be expressed in terms of the Bremsstrahlung integral \cite{nordsieck}. ${\bf K}$ is the local momentum vector of the $b-t$ system. For more details one is referred to Refs. \cite{rc, shyam}. 

The only input in the present theory is the projectile bound-state wavefunction $\phi_a^{\ell m}({\bf r}_1)$, which can be written as $\phi_a^{\ell m}({\bf r}_1) = i^{\ell} u_{\ell}(r)Y_{\ell\,m}(\hat{\bf r})$, where $u_{\ell}(r)$ is its radial part and $Y_{\ell\,m}(\hat{\bf r})$ are the spherical harmonics. To obtain the realistic radial wavefunction $u_{\ell}(r)$, I solve the radial Schr\"odinger equation with a Woods-Saxon potential, where the depth of the potential ($V_0$) is adjusted to get the binding energy of the projectile in a particular bound state.  
%%%%%%%%%%%%%%%%%%%%%%%%%%%%%%%%%%%%%%%%%%%%%%%%%%%%%%%%%%%%%%%%%%%%

\section{Results and discussions}
%\begin{widetext}
\begin{table*}[ht]
\begin{center}
\caption{FWHM calculated from the PMD of the core fragment in the Coulomb breakup of a projectile (considered in its different states with excitation energies E$_x$), on a Pb target at 100 MeV/nucleon beam energy.}
\label{ta1}

\begin{tabular}{c|c|c|c|c|c|c|c}

\hline\hline
%\\[-2.0ex]
S. No. & Nucleus& $J^{\pi}$    & E$_x$ (MeV)&  Single-particle configuration & $S_n$ (MeV) & $V_0$ (MeV) & FWHM (MeV/$c$)   \\
               
\hline
1. & $^{11}$Be&$\frac{1}{2}^+$& 0.0        & $^{10}$Be(0$^+$) $\otimes$ 2$s_{1/2}\nu$ &  0.501 & -71.03 & 43.23  \\
   &          &$\frac{1}{2}^-$& 0.320      & $^{10}$Be(0$^+$) $\otimes$ 1$p_{1/2}\nu$ &  0.182 & -43.58 & 41.15  \\
\hline
   &          & $1^+$         & 0.0        & $^{11}$B($\frac{3}{2}^-$) $\otimes$ 1$p_{1/2}\nu$ &  3.369 & -39.55 & 148.29  \\

   &          & $2^+$         & 0.953      & $^{11}$B($\frac{3}{2}^-$) $\otimes$ 1$p_{1/2}\nu$ &  2.417 & -30.88 & 128.37  \\
2. & $^{12}$B & $2^-$         & 1.673      & $^{11}$B($\frac{3}{2}^-$) $\otimes$ 2$s_{1/2}\nu$ &  1.697 & -49.65 & 68.36  \\
   &          & $1^-$         & 2.620      & $^{11}$B($\frac{3}{2}^-$) $\otimes$ 2$s_{1/2}\nu$ &  0.75 & -53.46 & 49.96  \\
   &          & $0^+$         & 2.723      & $^{11}$B($\frac{3}{2}^-$) $\otimes$ 1$p_{3/2}\nu$ &  0.647 & -36.17 & 68.64  \\
\hline
   &          & $\frac{1}{2}^-$& 0.0       & $^{12}$C(0$^+$) $\otimes$ 1$p_{1/2}\nu$ &  4.95 & -44.24 & 169.37  \\
3. & $^{13}$C & $\frac{1}{2}^+$& 3.089     & $^{12}$C(0$^+$) $\otimes$ 2$s_{1/2}\nu$ &  1.86 & -59.29 & 73.76  \\
   &          & $\frac{3}{2}^-$& 3.684     & $^{12}$C(0$^+$) $\otimes$ 1$p_{3/2}\nu$ &  1.27 & -30.26 & 97.34  \\
   &          & $\frac{5}{2}^+$& 3.853     & $^{12}$C(0$^+$) $\otimes$ 1$d_{5/2}\nu$ &  1.09 & -57.12 & 123.65  \\

\hline
\hline
\end{tabular}
\end{center}
\end{table*}
%\end{widetext}

Using the formalism given in the previous section, I calculate the parallel momentum distribution of the core fragment in the Coulomb breakup of a projectile on Pb at 100 MeV/nucleon. The nuclei I consider as projectiles are $^{10}$Be, $^{11}$Be, $^{12}$B, and, $^{13}$C in their different bound excited states. As mentioned in Sec. I, all these nuclei are suggested to have a one-neutron halo structure in their one or more excited states. Note that $^{11}$Be is also a well-known example of a ground-state one-neutron halo nucleus. I do not consider $^{9}$Be, in my study because the suggested halo state (1/2$^+$ at $E_x = 1.684$ MeV) is an {\it unbound state} and lies at 20 keV above the neutron-emission threshold. My intention is to look for the FWHM of the PMD of the core fragment in the breakup of the above mentioned nuclei in their different excited states, especially for the states that are predicted to have a halo structure. At this point, it is worth mentioning that the width of the PMD does not depend on the reaction mechanism \cite{shyam01}. Also it has been found in Refs. \cite{kidd,mermaz} that the width of the PMD remains nearly constant for a wide beam energy range (50 MeV/nucleon to 2 GeV/nucleon). 
Furthermore, many theoretical and experimental studies involving fragmentation reactions show that the width of the PMD does not depend on the target mass at all \cite{goldhaber,morrissey,sauvan,orr,meierbachtol,bertulani,banerjee}. 
With this background, I now start discussing all these cases in detail.

\begin{table*}[ht]
\begin{center}
\caption{FWHM from the PMD of $^9$Be in the Coulomb breakup of $^{10}$Be in its different excited states (having excitation energies E$_x$), on a Pb target at 100 MeV/nucleon beam energy. For a given $J^{\pi}$, the PMD is calculated by summing up the contributions from individual single-particle configurations considered here multiplied by their respective spectroscopic factors (S.F.), which are taken from Ref. \cite{khalili}.} 

\label{ta2}

\begin{tabular}{c|c|c|c|c|c|c}

\hline\hline
%\\[-2.0ex]
$J^{\pi}$ & E$_x$ (MeV) &  Single-particle configuration           & $S_n$ (MeV) & $V_0$ (MeV) & S.F. \cite{khalili}  & FWHM (MeV/$c$)   \\
               
\hline
$0^+$     &  0.0        & $^{9}$Be(3/2$^-$) $\otimes$ 1$p_{3/2}\nu$ &   6.812 & -53.99 &2.26          & 191.15  \\
\hline
$2^+$     &  3.368      & $^{9}$Be(3/2$^-$) $\otimes$ 1$p_{3/2}\nu$ &   3.444 & -43.98 &0.24          &  173.07 \\
          &             & $^{9}$Be(5/2$^-$) $\otimes$ 1$p_{3/2}\nu$ &   5.873 & -49.28 &1.17          &   \\
\hline
$2^+$     &  5.958      & $^{9}$Be(3/2$^-$) $\otimes$ 1$p_{3/2}\nu$ &   0.854 & -37.21 &0.28          &   \\
          &             & $^{9}$Be(3/2$^-$) $\otimes$ 1$p_{1/2}\nu$ &   0.854 & -46.31 &0.54          &  82.65 \\
          &             & $^{9}$Be(5/2$^-$) $\otimes$ 1$p_{3/2}\nu$ &   3.283 & -43.60 &0.23          &   \\
          &             & $^{9}$Be(5/2$^-$) $\otimes$ 1$p_{1/2}\nu$ &   3.283 & -52.29 &0.13          &   \\
\hline
$2^-$     &  6.263      & $^{9}$Be(3/2$^-$) $\otimes$ 2$s_{1/2}\nu$ &   0.549 & -72.00 &0.70          &   \\
          &             & $^{9}$Be(3/2$^-$) $\otimes$ 1$d_{5/2}\nu$ &   0.549 & -76.01 &0.16          &  44.65 \\
          &             & $^{9}$Be(5/2$^-$) $\otimes$ 2$s_{1/2}\nu$ &   2.978 & -82.59 &0.02          &   \\
          &             & $^{9}$Be(5/2$^-$) $\otimes$ 1$d_{5/2}\nu$ &   2.978 & -82.01 &0.10          &   \\

\hline
\hline
\end{tabular}
\end{center}
\end{table*}
%\end{widetext}

As a first case I consider the nucleus of $^{11}$Be, which has only two bound states and both have dominant single-particle configurations \cite{cappuzzello,aumann,liu2}. The ground state, which has an intruder $sd$ shell configuration, has spin-parity $J^\pi$ = 1/2$^+$, whereas the spin-parity $J^\pi$ of the first excited state is 1/2$^-$. These states are formed by coupling the 2$s_{1/2}$ and 1$p_{1/2}$ neutrons with 0$^+$ ground state of $^{10}$Be with one-neutron removal energy values ($S_n$) of 0.501 MeV and 0.182 MeV, respectively. Using the radius ($r_0$) and diffuseness ($a_0$) parameters of the Woods-Saxon potential as 1.15 fm and 0.5 fm \cite{shub3}, the potential depths required to reproduce binding energies of ground and first excited states are -71.03 MeV and -43.58 MeV, respectively. From the present calculations, the FWHM of the PMD of the $^{10}$Be core, obtained for the ground state, which is a well-known example of a halo, is 43.25 MeV/$c$ (also reported in Ref. \cite{shub3}), and agrees very well with the experimental value of $43.6 \pm 1.1$ MeV/$c$ \cite{kelley}. Interestingly, for the 1/2$^-$ excited state, I also get almost the same value of the FWHM, which is around 41.15 MeV/c. The narrow momentum distribution obtained for the first excited state in my calculations, therefore, indicates the possibility of halo formation in this state and hence favors the findings of Ref. \cite{belyaeva}.

As a second case I consider the $^{12}$B nucleus in its five low-lying states with $J^\pi$ of $1^+$ (E$_x$ = 0 MeV) , $2^+$ (E$_x$ = 0.953 MeV), $2^-$ (E$_x$ = 1.673 MeV), $1^-$ (E$_x$ = 2.620 MeV), and $0^+$ (E$_x$ = 2.723 MeV) and having $S_n$ values of 3.369, 2.417, 1.697, 0.75, and 0.647 MeV, respectively. Similar to Refs. \cite{liu,lin}, I consider the single-particle configurations, where states $1^+$, $2^+$ and states $2^-$, $1^-$ are obtained by coupling the $^{11}$B(3/2$^-$) ground state with $1p_{1/2}$ and $2s_{1/2}$ neutrons, respectively, and for the fourth excited state $0^+$, I consider the coupling of $^{11}$B(3/2$^-$) with a $1p_{3/2}$ neutron. With the radius and diffuseness parameters of the potential taken as 1.25 fm and 0.65 fm \cite{liu}, respectively, the values of $V_0$ for $1^+$, $2^+$, $2^-$, $1^-$, and $0^+$ states are -39.55, -30.88, -49.65, -53.46, -36.17 MeV, respectively. From the calculated PMD of the $^{11}$B core I found that the value of the FWHM is relatively much smaller for the second (2$^-$), third (1$^-$), and fourth (0$^+$) excited excited states. Therefore, these three states clearly show the signatures of a halo structure, and hence, my calculations favor the findings of Refs. \cite{liu,lin} where the (2$^-$) and (1$^-$) states of $^{12}$B are suggested to have possible halo structures because of the large radii obtained for these states. 

Next I consider the $^{13}$C nucleus and take into account its four low-lying states $1/2^-$ (E$_x$ = 0.0 MeV), $1/2^+$ (E$_x$ = 3.089 MeV), $3/2^-$ (E$_x$ = 3.684 MeV), and $5/2^+$ (E$_x$ = 3.853 MeV). I consider the same single-particle configuration as in Ref. \cite{shub4}, where these $1/2^-$, $1/2^+$, $3/2^-$, and $5/2^+$ states are constructed by coupling the $^{12}$C(0$^+$) ground state with the neutrons in the $1p_{1/2}$, $2s_{1/2}$, $1p_{3/2}$, and $1d_{5/2}$ orbitals, respectively. The Woods-Saxon parameters $r_0$ and $a_0$ are taken as 1.236 and 0.62 fm \cite{shub4}, respectively, and the values of $V_0$ required to reproduce the binding energies of ground, first, second and third excited states are -44.24, -59.29, -30.26, and -57.12 MeV, respectively.
In fact, $^{13}$C is very important from an astrophysical point of view \cite{wiescher,straniero} and the neutron capture reaction $^{12}$C($n, \gamma$)$^{13}$C is one of the processes by which it is formed. The total capture cross section is contributed by the capture to these four low-lying states \cite{shub4,ohsaki} considered here. Because of the large capture cross section, the first $1/2^+$ state of $^{13}$C was suggested to be a halo \cite{otsuka}, which also has an  extended density distribution \cite{otsuka1} and a large radius \cite{liu}. In the present PMD calculations, I also got a relatively narrower momentum distribution for this state as compared to the other three states. However, the associated FWHM (73 MeV/$c$) is somewhat larger than those of the well-established halo nuclei. In Table \ref{ta1}, I summarize my results for $^{11}$Be, $^{12}$B, and $^{13}$C.

As a last case I consider the nucleus $^{10}$Be in its four different bound states as considered in Ref. \cite{khalili} with the same single-particle configurations. $^{10}$Be is considered as the most stable isotope of Be with $S_n$ = 6.812 MeV, and is an example of the $N = 6$ magic number \cite{shub4,otsuka3}. It has been mentioned in Refs. \cite{khalili, arai2} that the 0$^+$ ground  state and first 2$^+$ (E$_x$ = 3.368 MeV) and 2$^-$ (E$_x$ = 6.263 MeV) excited states have reasonably good shell-model-like structure, whereas the other low-lying bound states 2$^+$ (E$_x$ = 5.958 MeV), 1$^-$ (E$_x$ = 5.959 MeV), and 0$^+$ (E$_x$ = 6.179 MeV) exhibit molecular structure. Among these excited states, 1$^-$ and 2$^-$ states were speculated to have halo structures in Ref. \cite{negoita}, however, because of the observed molecular structure the possibility of a halo structure of 1$^-$ was discarded in Ref. \cite{khalili}. For the present study, I follow Ref. \cite{khalili} for the single-particle configurations and respective spectroscopic factors. The Woods-Saxon parameters $r_0$ and $a_0$ in this case are same as those taken for $^{11}$Be. Additionally, I also take a spin-orbit strength of -10 MeV in this case to take care of different $j$ contributions corresponding to a given $\ell$. 
Table \ref{ta2}, presents the values of FWHM calculated from the PMD of the $^9$Be in the Coulomb breakup of $^{10}$Be in its different excited states, which are mixtures of various single-particle configurations \cite{khalili}.

 These different configurations are constructed by coupling the valence neutron in $s$ or $p$ or $d$ orbitals with the  $^9$Be core either in the ground state (3/2$^-$) or in the excited state (5/2$^-$) and by adjusting the potential depths (given in Table \ref{ta2}) to reproduce the corresponding neutron removal energies. For the configurations that involve the 5/2$^-$ excited state of the core, the total neutron removal energy is obtained by summing up the excitation energy of the 5/2$^-$ core (2.429 MeV) with the $S_n$ value when the core is in the ground state. The total parallel momentum distribution for a given bound state, is calculated by summing up the contributions of individual single-particle configurations multiplied by their respective spectroscopic factors, which are taken from Ref. \cite{khalili}. From the table it is clear that the $2^-$ state has the smallest value of FWHM (45 MeV/$c$) as compared to the others states considered here. This value (45 MeV/$c$) remains unchanged if I consider only the configuration where the 2$s_{1/2}$ neutron is coupled with the ground state of the $^9$Be core, because this is the dominant configuration and also the PMD of this configuration has a large magnitude as compared to the other three configurations of the $2^-$ state. The small value of the FWHM calculated for the $2^-$ state, therefore, suggest a halo structure and favors the predictions of Ref. \cite{khalili}, where the halo structure for this state was suggested by examining the electromagnetic transitions.

\section{Conclusions}
In conclusion, by using a fully quantum mechanical post-form FRDWBA theory, I have studied the Coulomb breakup of $^{10}$Be, $^{11}$Be, $^{12}$B, and $^{13}$C on a Pb target at 100 MeV/nucleon. The reason I use high beam energies is that at higher beam energies the higher order effects and postacceleration effects are negligible \cite{prabir}. From the calculated PMD of the core fragment, a specific reaction observable, I looked for the possibilities of halo structure in the different excited states of the projectiles considered in this study. In the present calculations a relatively narrow momentum distribution is obtained for the $2^-$ (1.673 MeV),  $1^-$ (2.620 MeV), $0^+$ (2.723 MeV) states of $^{12}$B, the $1/2^-$ (0.320 MeV) state of $^{11}$Be, the $2^-$ (6.263 MeV) state of $^{10}$Be and the $1/2^+$ (3.089 MeV) state of $^{13}$C. 
In some cases the calculated FWHM values are almost the same as those observed for the well-known ground-state one-neutron halo nuclei like $^{11}$Be and $^{19}$C. For  the $1/2^+$ state of $^{13}$C, the FWHM value is somewhat larger (around 74 MeV/$c$) but still it is much smaller as compared to all other low-lying states of it considered here. A narrow momentum distribution, therefore indicates the possibilities of a one-neutron halo structure in these states. A variation of 10\%-15\% in the potential parameters $r_0$ and $a_0$, results in only 1\%-2\% change in the present calculated FWHM values.   

I have therefore, used a reaction observable (PMD) in my calculations to predict the halo structure in the excited states of nuclei. The predictions of the present model agree with those from the other works that were based on the indirect radii measurements and electromagnetic transitions. This approach can further be used to predict more such cases. The new classes of experiments, such as the one reported in Ref. \cite{kobayashi2} to determine the interaction cross sections in the excited states, could be useful to confirm such predictions. Nevertheless, for the ground states where the experimental data exist, the calculations of the present theory agree with the data, for example, in Ref. \cite{shub3}, the FWHM values from our calculations agree with the experimental values for the cases of $^{11}$Be and $^{12}$Be.

In the past, the PMD has also been used to study the excited states of the core through the breakup or knockout reactions of the projectile \cite{shyam01,aumann,baumann}, but in the present study, to the best of my knowledge, for the first time I have used the PMD to study the excited states of the projectile for their halo structures.

%and the examine the possibility of halo structure in the excited state of the projectile. 
%We reiterate that such a study is possible 
%Although the Coulomb breakup experiments in the excited states are difficult but efforts are started in this directions \cite{fleurot}. 

\section*{Acknowledgment}
 The author acknowledges support from the U.S. NSF under Grant No. 1415656 and the U.S. DOE under Grant No. DE-FG02-08ER41533 and also thanks Rajdeep Chatterjee for useful correspondence.

\end{document}